\documentclass[runningheads,a4paper]{llncs}

\usepackage{amssymb}
\usepackage{graphicx}
\usepackage{subfigure}
\usepackage{listings}

\usepackage{amssymb}
\usepackage{amsmath}
\usepackage{url}
\usepackage{color}
\usepackage{hyperref}

\definecolor{dgreen}{rgb}{0.8,0,0}
\hypersetup{colorlinks=true,urlcolor=dgreen,linkcolor=dgreen,citecolor=dgreen}

\newcommand{\keywords}[1]{\par\addvspace\baselineskip
\noindent\keywordname\enspace\ignorespaces#1}

\begin{document}

\mainmatter

\title{How China Is Blocking Tor}

\author{Philipp Winter \and Stefan Lindskog}

\titlerunning{How China Is Blocking Tor}

\authorrunning{Philipp Winter and Stefan Lindskog}

\institute{
	Karlstad University \\
	\texttt{\{\href{mailto:philwint@kau.se}{philwint},\href{mailto:steflind@kau.se}{steflind}\}@kau.se}
}

\maketitle

\begin{abstract}
Not only the free web is victim to China's excessive censorship, but also the Tor anonymity network:
the Great Firewall of China prevents thousands of potential Tor users from accessing the network.

In this paper, we investigate how the blocking mechanism is implemented, we conjecture how China's
Tor blocking infrastructure is designed and we propose countermeasures.

Our work bolsters the understanding of China's censorship capabilities and thus paves the way
towards more effective evasion techniques.
\end{abstract}

\section{Introduction}
\label{sec:introduction}
On October 4, 2011 a user reported to the Tor bug tracker that unpublished bridges stop working
after only a few minutes when used from within China \cite{ticket1}. Bridges are unpublished Tor
relays and their very purpose is to help censored users to access the Tor network if the ``main
entrance'' is blocked. The bug report indicated that the Great Firewall of China (GFC) has been
enhanced with the potentiality of dynamically blocking Tor.

This censorship attempt is by no means China's first effort to block Tor. In the past, there have
been efforts to block the website \cite{tpo2}, the public Tor network \cite{tpo1,tpo3} and parts of
bridges \cite{tpo4}. According to reports, these blocks were realised by simple IP blacklisting
and HTTP header filtering \cite{tpo2}. All these blocking attempts had in common that
they were straightforward and inflexible.

In contrast to the above mentioned censorship attempts, the currently observable block appears to be
much more flexible and sophisticated. Freshly set up bridges get blocked within only minutes. In
addition, the GFC blocks bridges dynamically without simply enumerating their IP addresses and
blacklisting them (cf. \cite{Ling2012}).

In this paper, we try to deepen the understanding of the infrastructure used by the GFC to block the
Tor anonymity network. Our contributions are threefold:
\begin{enumerate}
	\item We reveal how Chinese users are hindered from accessing the Tor network.
	\item We conjecture how China's Tor blocking infrastructure is designed.
	\item We discuss evasion strategies.
\end{enumerate}

The remainder of this paper is structured as follows. Section \ref{sec:related} gives an overview of
previous work. Section \ref{sec:setup} introduces our experimental setup. Section \ref{sec:results}
presents how users are prevented from using Tor and Section \ref{sec:analysis} analyses China's
underlying infrastructure. Evasion strategies are discussed in Section \ref{sec:evasion}. Finally,
the paper is concluded in Section \ref{sec:conclusion}.

\section{Related Work}
\label{sec:related}
Our work is not the first documentation of the current Chinese attempts to block the Tor anonymity
network. In \cite{twilde}, Wilde revealed first crucial insights about the block. Over a period of
one week in December 2011, Wilde analysed how \emph{unpublished} Tor bridges are getting scanned
and, as a result, blocked by the GFC. Wilde's experiments relied upon a virtual private server (VPS)
inside China and several Tor EC2 cloud bridges \cite{ec2,TorCloud} located in the USA.

\begin{figure}
\centering
\includegraphics[width=0.7\textwidth]{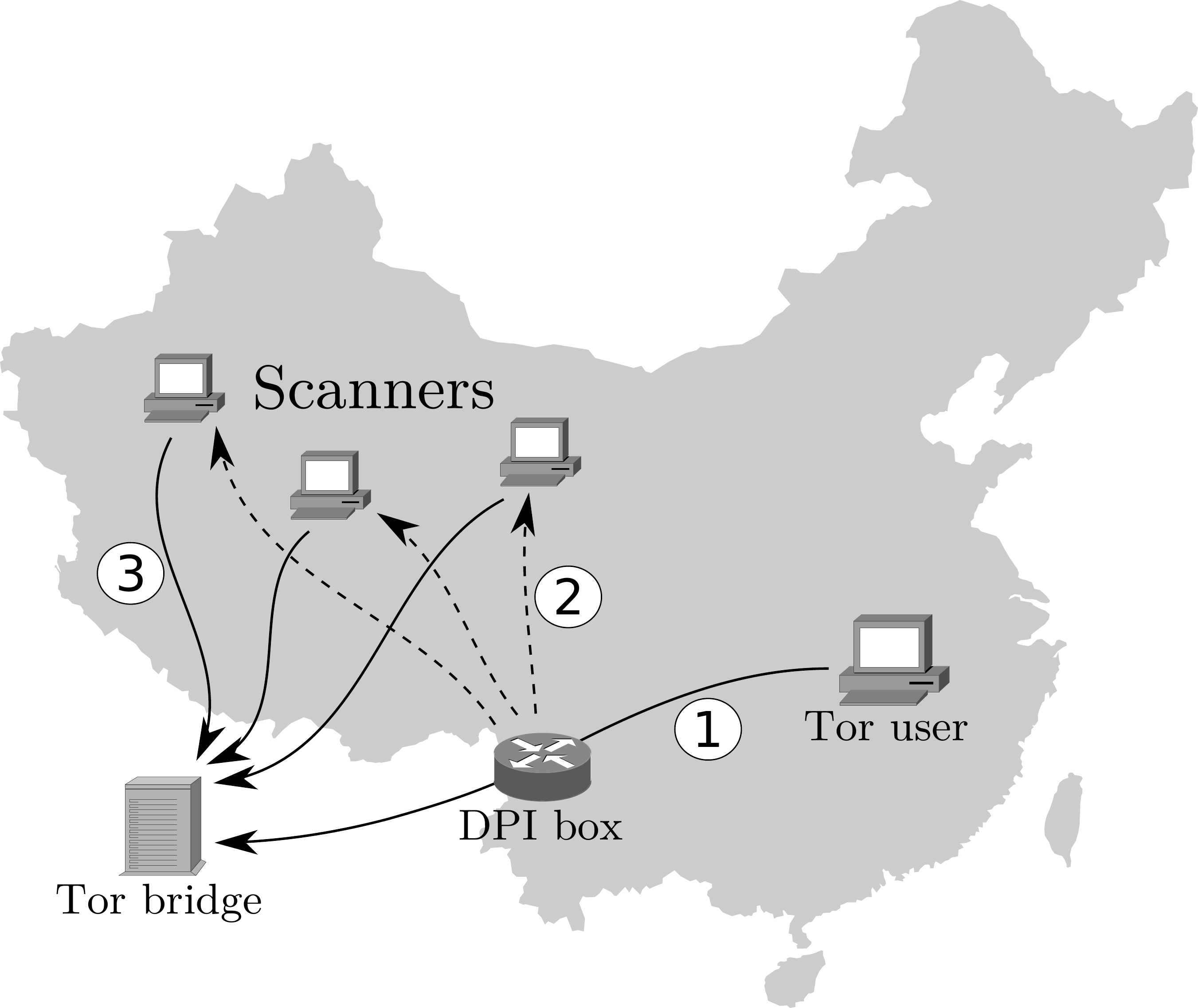}
\caption{The structure of the Chinese Tor blocking infrastructure. After DPI boxes identified a Tor
connection to a bridge or relay, active scanners connect to the same machine and induce the block if
the machine ``speaks Tor''.}
\label{fig:china}
\end{figure}

According to Wilde's results, the basic functionality of the Chinese Tor blocking infrastructure is
depicted in Figure \ref{fig:china}. When a Tor user in China establishes a connection to a bridge or
relay, deep packet inspection (DPI) boxes \emph{recognise} the Tor protocol (1). Shortly after a Tor
connection is detected, \emph{active scanning} is initiated (2). The scanning is done by seemingly
random Chinese IP addresses. The scanners connect to the respective bridge and try to
\emph{establish a Tor connection} (3). If it succeeds, the bridge is \emph{blocked}.

Wilde was able to narrow down the suspected cause for active scanning to the cipher list sent by the
Tor client inside the TLS client hello\footnote{The TLS client hello is sent by the client after a
TCP connection has been established. Details can be found in the Tor design paper
\cite{Dingledine2004}.}. This cipher list appears to be unique and only used by Tor. That gives the
GFC the opportunity to easily identify Tor connections. Furthermore, Wilde noticed that active
scanning is done at multiples of 15 minutes. The GFC launches several scanners to connect to the
bridge at the next full 15 minute multiple when a Tor cipher list was detected. An analysis of the
Tor debug logs yielded that Chinese scanners initiate a TLS connection, conduct a renegotiation and
start building a Tor circuit, once the TLS connection was set up. After the scan succeeded, the IP
address together with the associated port (we now refer to this as ``IP:port tuple'') of the freshly
scanned bridge is blocked resulting in Chinese users not being able to use the bridge anymore.

In addition to the above mentioned key findings, Wilde provided additional data to strengthen his
results such as traceroutes to suspected scanners, a sample of IP addresses believed to be scanners
as well as an analysis of what is known as \emph{garbage scanning}: Chinese scanners sending
seemingly random binary data to TLS-speaking hosts. The garbage scanning does not seem to be related
to Tor although it seems to make use of the same infrastructure.

With respect to Wilde's contribution, we (a) revisit many experiments in greater detail and with
significantly more data, we (b) rectify observations which we did not find to be true anymore and we
(c) answer yet open questions.

\section{Experimental Setup}
\label{sec:setup}
We begin this section by discussing the legal and ethical aspects of our work. We then discuss the
vantage points we used in our analysis and point out the shortcomings.

\subsection{Legal and Ethical Aspects}
During the process of preparing and running our experiments we took special care to not violate any
laws. In addition, all our experiments were in accordance with the terms of service of our service
providers. In order to ensure reproducibility and encourage further research, we publish our
gathered data and developed code\footnote{Both can be found at:
\url{http://www.cs.kau.se/philwint/static/gfc/}.}. The data includes IP addresses of Chinese hosts
which were found to conduct active scanning of our Tor bridge. We carefully configured our Tor
bridges to remain unpublished and we always picked randomly chosen high ports to listen to so we can
say that the data is free from legitimate Tor users.

\subsection{Vantage Points}
In order to ensure a high degree of confidence in our results, we used different vantage points. We
had a relay in Russia, bridges in Singapore and Sweden and clients in China.

\textbf{Bridge in Singapore}: A large part of our experiments was conducted with our Tor bridge
located in Singapore. The bridge was running inside the Amazon EC2 cloud \cite{ec2}. The OpenNet
Initiative reports Singapore as a country conducting minimal Internet filtering involving only
pornography \cite{opennet}. Hence, we assume that our experimental results were not falsified by
Internet filtering in Singapore.

\textbf{Bridge in Sweden}: To reproduce certain experiments, we set up Tor bridges located at our
institution in Sweden. Internet filtering at Karlstad University is limited to well-known malware
ports which we avoided during the experiments. Thus, we can rule out filtering mechanisms
interfering with our results.

\textbf{Clients in China}: To avoid biased results, we used two types of vantage points inside
China: open SOCKS proxies and a VPS.

We compiled a list of public Chinese SOCKS proxies by searching Google. We were able to find a total
of 32 SOCKS proxies which were distributed amongst 12 distinct autonomous systems. We used the
SOCKS proxies to rerun certain experiments on a smaller scale to rule out phenomena limited to our
VPS.

Our second vantage point and primary experimental machine is a rented VPS. The VPS ran Linux and
resided in the autonomous system number (ASN) 4808. We had full root access to the VPS which made it
possible for us to sniff traffic and conduct experiments below the application layer such as packet
fragmentation and manipulation.  Most of our experiments were conducted from our VPS whereas the
SOCKS proxies' primary use was to verify results.

\textbf{Relay in Russia}: A public Tor relay located in a Russian data center was used to
investigate the blocking public Tor relays are undergoing. The relay served as middle relay, meaning
that it does not have direct contact with Tor users and it does not see their exit traffic.

\subsection{Shortcomings}
Heisenberg's \emph{uncertainty principle} is not limited to physics \cite{Jia2009} but also applies
to active network measurement to some extent. Injecting packets into a network for the sake of
measuring a phenomena inevitably interferes with the observed results. This is particularly critical
when measuring censorship systems.

Active analysis of a censorship system can easily attract the censor's attention if no special care
is taken to ``stay under the radar''. Due to the fact that China is a sophisticated censor with the
potential power to actively falsify measurement results, we have to point out potential shortcomings
in our experimental setup.

First of all, we have no reliable information about the owners of our public SOCKS proxies.
Whois-lookups did not yield anything suspicious but the information in the records can be spoofed.
Second, our VPS was located in a data center where Tor connections typically do not originate.
We also had no information about whether our service provider conducts Internet filtering and
the type or extent thereof.

We are aware of the above mentioned shortcomings and designed our experiments in a way to minimise
falsifying effects. We reproduced our experiments from different vantage points to eliminate
phenomena limited to one point of view.

\section{Why Alice Cannot Use Tor in China}
\label{sec:results}
This section shows how the Tor anonymity network is blocked in China. We focus on the blocking
mechanisms. A detailed analysis of how the blocking is designed is provided in Section
\ref{sec:analysis}. We present our findings by accompanying Alice who lives in China and wants to
use the Tor network. We point out all the difficulties she faces during her journey towards a free
Internet ranging from accessing the Tor website \cite{tpo0} to the use of bridges.

\subsection{Website Block}
First, the Tor application is required to use the anonymity network. One way to get the binary is to
download it from the website \cite{tpo0}. This is where the censorship begins because the website is
partially inaccessible in China. We discovered that the GFC monitors the ``Host'' field in the HTTP
header\footnote{Something similar has been observed before \cite{tpo2}.}. If the field equals
``torproject.org'', the connection is reset by injected RST segments as described in
\cite{Clayton2006}. Slight variations of the field's content, such as ``orproject.org'' and
``torproject.or'' do not lead to injected RST segments.

We tracked down the cause of the injected RST segments to the ``Host'' field by sending the
same HTTP request to different IP addresses. Our connection was also reset after connecting to
``innocent'' domains such as gmail.com, kau.se and baidu.com. In addition, we were able to connect
to one of the Tor Project's web servers and send a GET request for gmail.com. We received an error
message from the web server without any interference from the GFC. Furthermore, we received the
correct DNS reply for ``torproject.org'' from our Chinese DNS servers. We conclude that the
string ``torproject.org'' in the ``Host'' HTTP field causes injected RST segments and
that the IP addresses of the Tor Project's web servers are not blacklisted.

We found mirrors whose hostname do not contain ``torproject.org'' to be reachable too. And since the
official Tor website offers HTTPS, all Alice has to do is to add an ``s'' after ``http'' in her
browser in order to make the official web site load.

\subsection{The Public Tor Network}
Now that Alice could download the Tor Browser Bundle, she immediately starts it. Without further
configuration, her client first tries to connect to the directory authorities and fetch the
consensus which contains the list of all public Tor relays.

However, all but one of the nine directory authorities are \emph{blocked on the IP layer}. From
within China they neither respond to pings, nor are any TCP services available. From our unfiltered
connection in Sweden we were able to contact the directory authorities over ICMP and TCP. We have no
explanation, however, why the directory authority with the IP address 212.112.245.170 is not
blocked.

If Alice is lucky, her client might connect to the unblocked directory authority and is then able to
download the consensus. But even then Alice has to face disappointment again. As we will describe
in Section \ref{sec:analysis} in detail, almost all of the public relays are blocked from inside
China.

From our middle relay in Russia we were able to verify that the blocking is realised by dropping the
SYN/ACK segment sent by the relay to the client. The SYN segment sent by Alice's client to the relay
passes the GFC unhindered.

\subsection{Bridges}
So far Alice could download the Tor client application but she was unable to use the public network.
The Tor Project came up with the concept of bridges \cite{Dingledine2006a} since it is not hard for
a censor to block access to the public network. Bridges are essentially relays which are not listed
in the public consensus. Their robustness relies on the assumption that it is hard for censors to
find out all their IP addresses and block them.

Alice does not give up and tries her luck with bridges. She obtains some bridge
descriptors\footnote{A bridge (and relay) descriptor is a set of information needed by the Tor
client to connect to the bridge.} from the bridge distribution website \cite{bridges}.
Unfortunately, none of the received bridges were reachable. As it turns out, they are blocked
\emph{in the same way} as public relays: Alice's client can send SYN segments to them and they reply
with a SYN/ACK segment, but the reply segment never reaches its intended destination.

\section{Analysis}
\label{sec:analysis}
The previous section revealed that the Tor network is blocked on at least \emph{three layers}
consisting of the website, the directory authorities and the relays (including bridges). In this
section we take a closer look at the blocking \emph{infrastructure}. In particular, we look at the
origin of the active scans.

\subsection{How Long Do Bridges Remain Blocked?}
So far it is unclear for \emph{how long} a freshly blocked bridge remains blocked. To answer this
question, we invoked two Tor processes on our bridge in Singapore. Before this experiment we changed
the IP address of the machine in Singapore in order to avoid interference with previous experiments.
Both Tor processes were private bridges and listening on TCP port 27418 and 23941. Both ports were
chosen randomly.

In the next step, we made the GFC block both IP:port tuples by initiating Tor connections to them
from our VPS in China. After the GFC started to block both tuples, we set up iptables
\cite{iptables} rules to whitelist our VPS in China to port 23941 and drop all other connections.
That way, the tuple should appear unreachable to the world with the exception of our Chinese VPS.
Port 27418 remained unchanged and hence reachable. We then started monitoring the reachability of
both Tor processes by continuously trying to connect to the processes using telnet from our VPS.

After approximately 12 hours, the Tor process behind port 23941 (which appeared to be unreachable to
the GFC) became reachable again whereas connections to port 27418 still timed out and continued to
do so. In our iptables logs we could find numerous connection attempts stemming from Chinese
scanners. This observation shows that once a Tor bridge has been blocked, it only remains
blocked if Chinese scanners are able to \emph{continuously connect} to the bridge. If they cannot,
the block is \emph{removed}.

\subsection{Is the Public Network Reachable?}
In a previous section we shortly mentioned that most of the public relays are blocked. To verify
this, we downloaded the published consensus from February 23 at 09:00. At the time, the
consensus contained descriptors for a total of 2819 relays. Then, from our Chinese VPS we tried to
establish a TCP connection to the Tor port of every single relay. If we were able to successfully
establish a TCP connection we classified the relay as reachable, otherwise unreachable.

We found that our Chinese client could successfully establish TCP connections with 47 out of all
2819 (1.6\%) public relays. We manually inspected the descriptors of the 47 relays, but could not
find any common property which could have been responsible for the relays being unblocked in China.
We checked the availability of the reachable relays again after a period of three days. Only
one out of the 47 unfiltered relays was still reachable.

\subsection{Where Does the Filtering Happen?}
We want to gain a better understanding of where the Chinese DPI boxes are looking for the Tor
fingerprint. In detail, we try to investigate whether the DPI boxes also analyse domestic and
ingress traffic.

We used six open Chinese SOCKS proxies distributed across four autonomous systems (ASN 4134, 4837,
9808 and 17968) to investigate domestic filtering. We simulated the initiation of a Tor connection
multiple times to randomly chosen TCP ports on our VPS, but could not attract any active scans.

Previous research confirmed that Chinese HTTP keyword filtering is bidirectional \cite{Clayton2006},
i.e., keywords are scanned in both directions. We wanted to find out whether this is true for the
Tor DPI infrastructure too. To verify that, we tried to initiate Tor connections to our Chinese VPS
from our vantage points in Sweden, Russia and Singapore. Despite multiple attempts we were not able
to attract a single scan.

The above mentioned results indicate that Tor filtering is probably only done at Chinese
\emph{border ASes} and only with traffic going from \emph{inside China to the outside world}. We
believe that there are two reasons for this. First, filtering domestic traffic in addition to
international traffic would dramatically increase the amounts of data to analyse since domestic
traffic is believed to be the largest fraction of Chinese traffic \cite{hroberts}. Second, at the
time of this writing there are no relays in China so there is no need to filter domestic or ingress
traffic. Tor usage in China means being able to connect to the outside world.

\subsection{What Pattern Is Matched by the GFC?}
Wilde discovered that the GFC identifies Tor connections by searching for the cipher list sent by
Tor clients \cite{twilde}. The cipher list is part of the TLS client hello which is sent by the Tor
user to the relay or bridge after a TCP connection has been established. This particular cipher
list, see Listing \ref{lst:cipherlist}, appears to be unique to Tor.

\begin{lstlisting}[basicstyle=\footnotesize\ttfamily,caption={Tor cipher list inside TLS client
hello.},label={lst:cipherlist},captionpos=b]
        c0 0a c0 14 00 39 00 38 c0 0f c0 05 00 35 c0 07
        c0 09 c0 11 c0 13 00 33 00 32 c0 0c c0 0e c0 02
        c0 04 00 04 00 05 00 2f c0 08 c0 12 00 16 00 13
        c0 0d c0 03 fe ff 00 0a 00 ff
\end{lstlisting}

According to our experiments, the Chinese DPI boxes are still looking for this particular cipher
list. However, we show that the mere occurence of the cipher list anywhere in the payload is not
enough to trigger scanning. The \emph{context} seems to be relevant.

In order to verify that the context is relevant, we embed the TLS handshake pattern inside an HTTP
request. The TLS handshake was disguised as User-Agent as can be seen in Listing \ref{lst:http}.

\begin{lstlisting}[basicstyle=\footnotesize\ttfamily,caption={Crafted HTTP
request.},label={lst:http},captionpos=b]
                   GET / HTTP/1.1
                   Host: baidu.com
                   User-Agent: <HANDSHAKE>
\end{lstlisting}

We proceeded to send this pseudo HTTP request to our bridge in Singapore. Despite several attempts
with slightly modified requests, we were not able to trigger scanning. We then set the first 6 bytes
in the same HTTP requests to \texttt{0x00} and sent it again. This time, we were able to trigger
scanning. We assume that the ``GET'' string at the beginning of the payload made the DPI boxes load
a different set of patterns to match.

\subsection{Where Are the Scanners Coming From?}
An important question we have not answered yet is where the active scans originate. To get extensive
data for answering this question, we continuously attracted active scanners over a period of 17 days
ranging from March 6 to March 23. We attracted scanners by simulating Tor connections from within
China to our bridge in Singapore\footnote{We repeated the same experiment with a bridge in Sweden
and with open SOCKS proxies and the findings were the same.}. For the simulation of Tor connections,
we created a program whose only purpose was to send the Tor TLS client hello to the bridge and
terminate afterwards. This is enough ``bait'' to attract active scanners. After every Tor connection
simulation, our program remained inactive for a randomly chosen value between 9 and 14 minutes. The
experiment yielded \emph{3295 scans of our bridge}. Our findings based on this data are described below.

\subsubsection{Scanner IP Address Distribution}
We are interested in the scanner's IP address distribution, i.e., how often can we find a particular
IP address in our logs? Our data exhibits two surprising characteristics:
\begin{enumerate}
	\item More than half of all connections---1680 of 3295 (51\%)---were initiated by \emph{a
	single} IP address: 202.108.181.70.
	\item The second half of all addresses is almost \emph{uniformly distributed}. Among all 1615
	remaining addresses, 1584 (98\%) were unique.
\end{enumerate}

The IP address 202.108.181.70 clearly stands out from all observed scanners. Aside from its heavy
activity we could not observe any other peculiarities in its scanning behaviour. The whois-record of
the IP address in Listing \ref{lst:mswhois} states a company named ``Beijing Guanda Technology
Co.Ltd'' as owner.

\begin{lstlisting}[basicstyle=\footnotesize\ttfamily,caption={Whois-record of the master
scanner.},label={lst:mswhois},
captionpos=b]
     inetnum:        202.108.181.0 - 202.108.181.255
     netname:        BJ-GD-TECH-CO
     descr:          Beijing Guanda Technology Co.Ltd
     country:        CN
     admin-c:        CH455-AP
     tech-c:         SY21-AP
     mnt-by:         MAINT-CNCGROUP-BJ
     changed:        suny@publicf.bta.net.cn 20020524
     status:         ASSIGNED NON-PORTABLE
     source:         APNIC

     [...]
\end{lstlisting}

We could only find a company named ``Guanda Technology Amusement Equipment Co., Ltd'' on the
Internet. It is not clear whether this is the same company. However, as explained below, we have
reason to believe that the scanner's IP addresses are spoofed by the GFC so the owner of the IP
address, assuming that it even exists, may not be aware of the scanning activity.

\subsubsection{Whois Records}
To learn more about the origin of the scanners we iterated over the entire corpus of gathered
scanner IP addresses and fetched the associated whois records. We manually investigated the obtained
records and discovered that all IP addresses are under the control of only two ISPs: \emph{China
Telecom} and \emph{China Unicom}\footnote{Many IP addresses were owned by CNCGroup (China Netcom
Group) as well, but this ISP is now merged with China Unicom.}. Both ISPs do not limit their scanner
IPs to one single network, but instead make use of networks distributed over multiple Chinese
provinces. Using our whois data we found out that the scanner IPs originated from 135 different
networks of varying size.

Judging by the number of subscribers, China Telecom and Unicom are the two biggest ISPs not just in
China, but also worldwide \cite{chinaISP}. In addition, both are run by the state. These two
prerequisites are possibly perfect breeding ground for censorship.

\subsubsection{Autonomous System Origin}
We used the IP to ASN mapping service provided by Team Cymru \cite{ip2asn} to get the autonomous
system number for every observed scanner. The result reveals that all scanners come from \emph{one
of three} ASes\footnote{Recent research efforts showed that China operates 177 autonomous systems
\cite{Roberts2011}.} with the respective percentage of observed scanners in parantheses:
\begin{itemize}
	\item \emph{AS4837}:\\CHINA169-BACKBONE CNCGROUP China169 Backbone (65.7\%)
	\item \emph{AS4134}:\\CHINANET-BACKBONE No.31,Jin-rong Street (30.5\%)
	\item \emph{AS17622}:\\CNCGROUP-GZ China Unicom Guangzhou network (3.8\%)
\end{itemize}

AS4134 is owned by China Telecom and AS4837 as well as AS17622 is owned by China Unicom. AS4134 and
AS4837 are the two largest ASes in China \cite{Roberts2011} and play a crucial role in the
country-wide censorship as pointed out by Xu, Mao and Halderman \cite{Xu2011}. Furthermore, Roberts
et al. \cite{Roberts2011} found out that China's AS level structure is \emph{far from uniform} with
a significant fraction of the countries traffic being routed through AS4134 or AS4837.

\subsubsection{Reverse DNS}
Using Google's open DNS server 8.8.8.8 we conducted reverse DNS lookups for all collected IP
addresses. This process yielded 284 DNS PTR records. With the exception of several misconfigured
records which were resolved to \emph{hn.kd.dhcp}, \emph{hn.kd.ny.adsl} and \emph{hn.kd.pix}, all
records contained either the substring \emph{dynamic} or \emph{adsl} which indicates that the IP
addresses originate from an ISP's address pool. All valid DNS PTR records ended with one of the
following four domain names:
\begin{itemize}
	\item \emph{163data.com.cn} (China Telecom)
	\item \emph{cndata.com} (China Telecom)
	\item \emph{sx.cn} (China Unicom Shanxi Province Network)
	\item \emph{jlccptt.net.cn} (China Unicom Jilin Province Network)
\end{itemize}

\subsubsection{IP Address Spoofing}
In his analysis, Wilde \cite{twilde} already tried to find evidence for a supposable IP address
hijacking conducted by the GFC. The idea was that the GFC might not have its own dedicated address
range---which would be trivial to block for Tor bridges and relays---but instead shortly hijacks
otherwise assigned IP addresses for the sole purpose of scanning. However, Wilde's gathered
traceroute data did not contain any indication for that hypothesis since the traceroute paths during
and after scanning did not differ.

During manual tests we noticed that sometimes shortly after a scan, the respective IP address starts
replying to ping requests\footnote{Note that this is never the case during or immediately after
scans. All ICMP packets are being dropped.}. In order to have more data to analyse, we wrote a
script to automatically gather additional data as soon as a scanner connects and again some minutes
afterwards. In particular, the script did the following:
\begin{itemize}
	\item Run TCP, UDP and ICMP traceroutes immediately after a scan and again 15 minutes later.
	\item Continuously ping the scanning IP address for 15 minutes.
	\item Capture all network traffic during these 15 minutes using tcpdump.
\end{itemize}

Between March 21 and 26 we started an independent round of attracting scanners and let our script
gather the above mentioned additional data. We caused a total of 429 scans coming from 427 unique IP
addresses. From all 429 scans we then extracted all connections where the continuous 15 minute ping
resulted in at least one ping reply packet. This process yielded 85 connections which corresponds to
approximately 20\% of all observed connections.

We analysed the 85 connections by computing:
\begin{itemize}
	\item The amount of minutes until a particular IP address started replying to our ping requests.
	\item The IP TTL difference (new ttl -- old ttl) between packets during the scan and ping
	replies.
\end{itemize}

\begin{figure}
\centering
\subfigure[The IP TTL difference between after and during the scan.]{
\includegraphics[scale=.45]{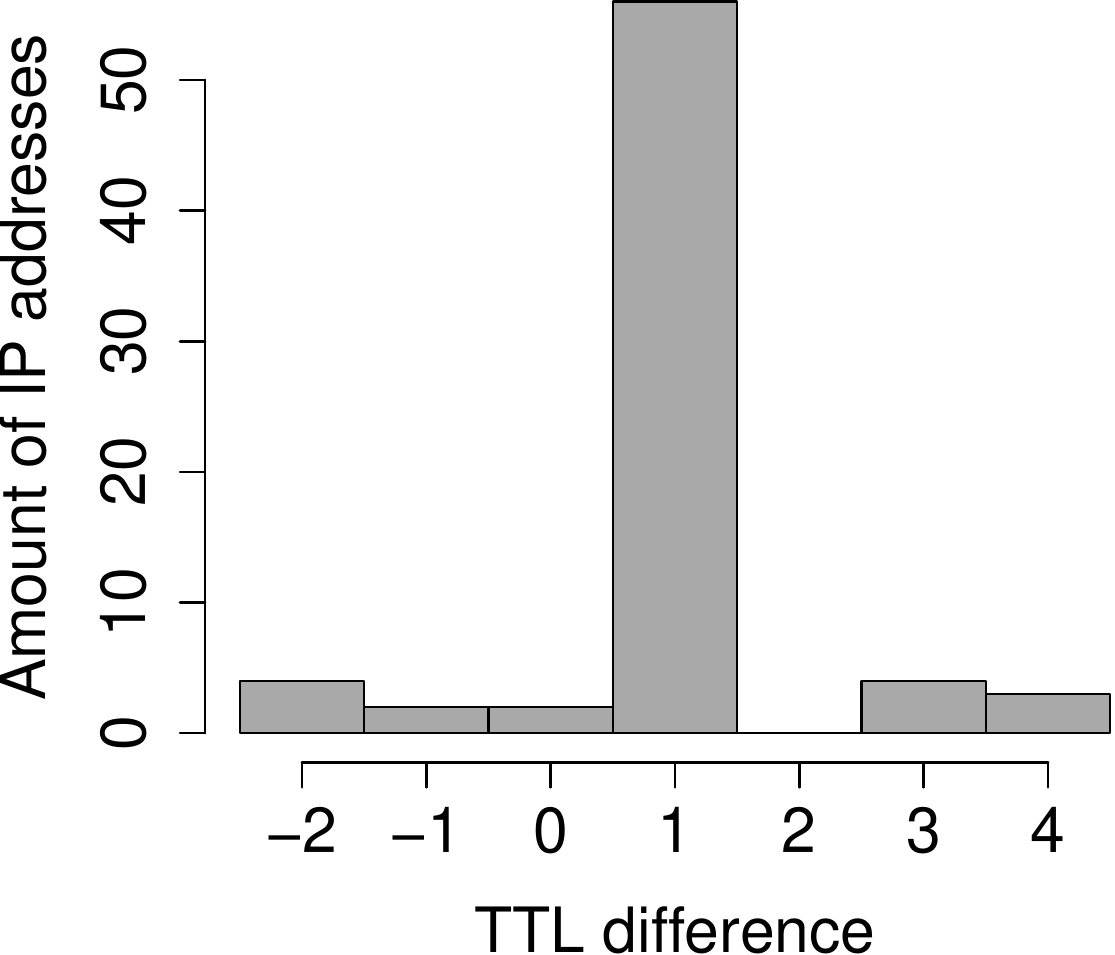}
\label{fig:subfig1}
}
\subfigure[The amount of minutes until the host started replying to pings.]{
\includegraphics[scale=.45]{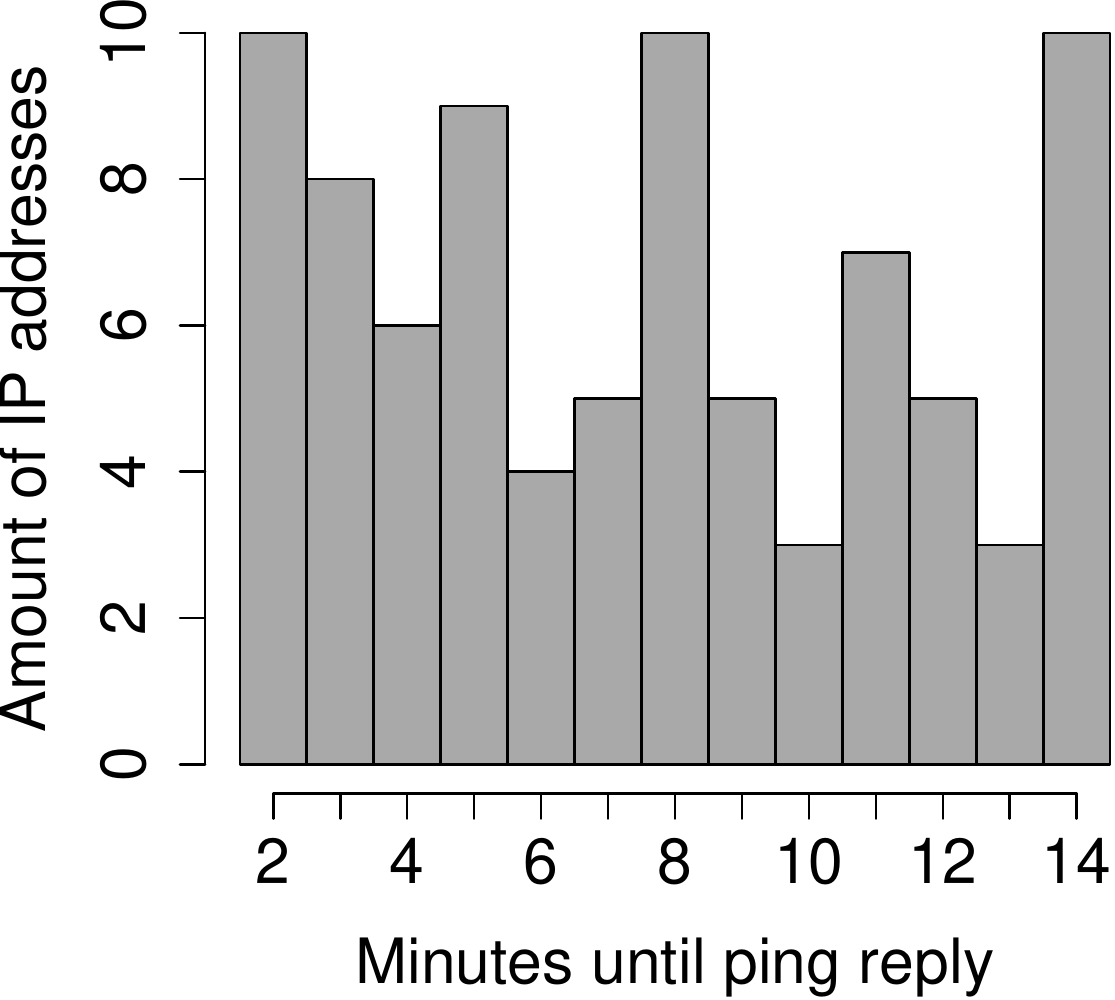}
\label{fig:subfig2}
}
\label{fig:subfigureExample}
\caption{Analysis of the IP TTL difference \subref{fig:subfig1} and the duration until ping replies
\subref{fig:subfig2}.}
\end{figure}

The results are illustrated in Figure \ref{fig:subfig1} and \ref{fig:subfig2}. Figure
\ref{fig:subfig1} depicts all IP TTL differences between after the scan (when the host starts
replying to ping packets) and during the scan. We had 14 outliers with a TTL difference of 65 and
192 but did not list them in the histogram. It is clearly noticeable that the difference was mostly
1, meaning that after the scan, the TTL was 1 higher than during the scan. Figure \ref{fig:subfig2}
illustrates how long it took for the hosts to start replying to the ping requests. No clear pattern
is visible.

An example of a traceroute is given in Listing \ref{lst:tr1} and \ref{lst:tr2}. We ran an ICMP
traceroute to the IP address 60.208.176.112 during and 15 minutes after the scan. While the
traceroute timed out during the scan, it finished successfully 15 minutes later. We point out that
the traceroute during the scan timed out \emph{just one hop} before the destination IP address.

One explanation for the changing TTL is that the GFC could be \emph{spoofing IP addresses}. The GFC
could be abusing several IP address pools intended for Internet users to allocate short-lived IP
addresses for the purpose of scanning. We further speculate that sometimes the GFC might be
spoofing IP addresses which are currently in use. When the spoofing ends, we are able to communicate
with the ``underlying'' host which seems to be 1 hop behind the scanning machine. This could mean
that the scanning infrastructure is located at data centers of ISPs.

However, we stress that the GFC conducting IP spoofing is a mere hypothesis. There might be
legitimate reasons for the observed TTL differences. From the GFC's point of view, spoofing ordinary
client IP addresses would be effective since bridges and relays can not block the scanners at
the IP layer. This would imply blocking legitimate Internet users from accessing the Tor network.

\subsection{When Do the Scanners Connect?}
In \cite{twilde}, Wilde stated that all scanners connected within three minutes after multiples of
15 minutes, i.e., during the minute intervals $[0,3]$, $[15,18]$, $[30,33]$ and $[45,48]$. While
these results were in accordance with our first manual tests during January, they did not hold true
for our scanner attraction experiments after January.

\begin{figure}
\centering
\includegraphics[width=0.75\textwidth]{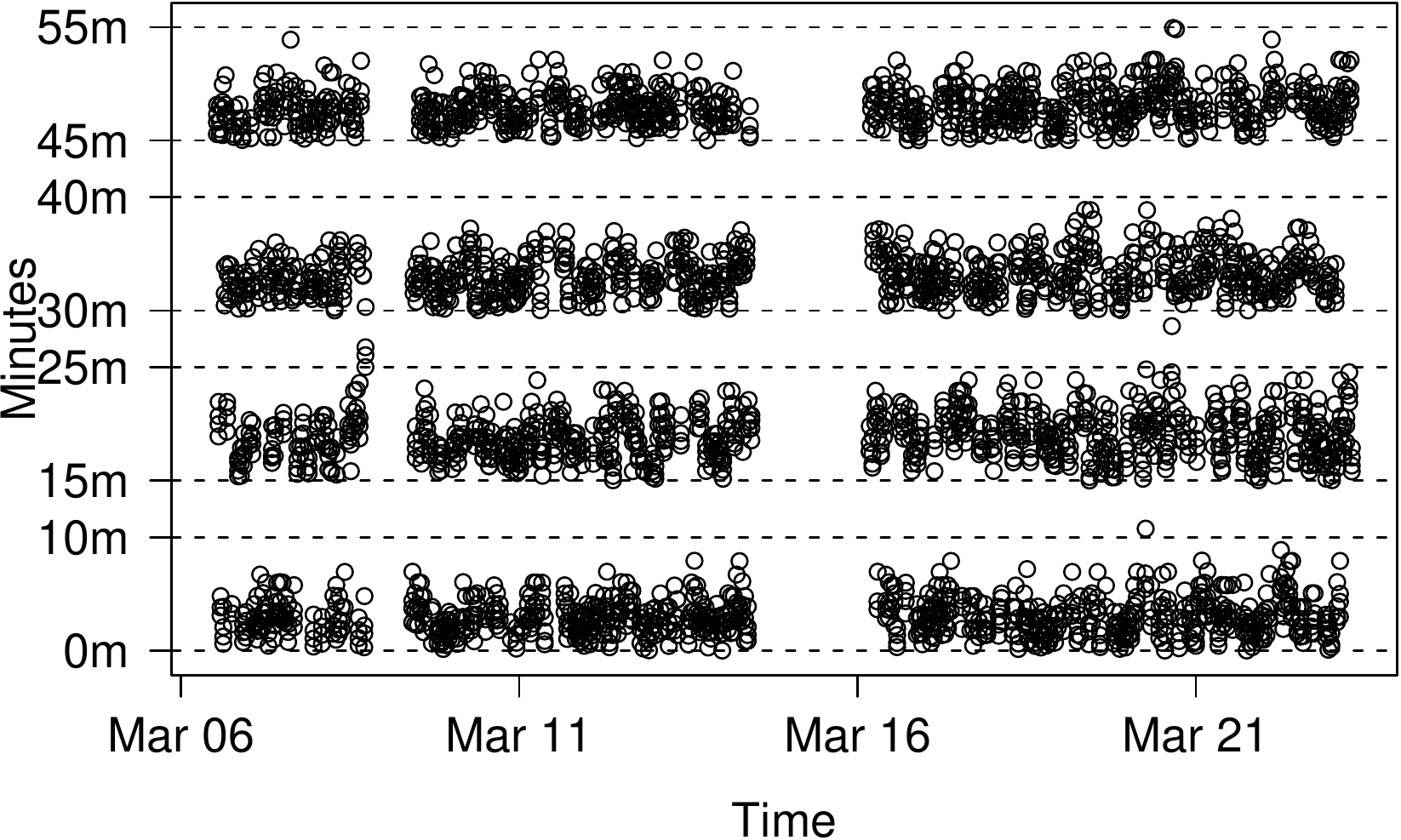}
\caption{Timing pattern of scanners connecting to our bridge. The $x$-axis depicts the time in days
whereas the $y$-axis represents the minute when a scanner connected to the bridge.}
\label{fig:timing}
\end{figure}

Figure \ref{fig:timing} visualises when scanners connect. The $y$-axis depicts the minutes of the
respective hour. Contrary to December 2011, when Wilde ran his experiments, the scanners now seem to
use a broader time interval to launch the scans. In addition, the data contains two time intervals
which are free from scanning. These intervals lasted from March 8 at around 17:30 to March 9 at
10:00 and from March 14 at 10:30 to March 16 at 4:30. We have no explanation why the GFC did not
conduct scanning during that time.

Closer manual analysis yielded that the data exhibits a \emph{diurnal pattern}. In order to make the
pattern visible, we processed the data as four distinct time series with every 15 minute interval
forming one time series, respectively. We smoothed the data points in the time series using simple
exponential smoothing as defined in Equation \ref{equ:ses}. The algorithm takes our data points $x$
as input and produces the smoothed time series $\hat{x}$.

\begin{equation}
\label{equ:ses}
\hat{x}_{t} =
\begin{cases}
x_{0} & \mbox{if } t \le 1 \\
\alpha \cdot x_{t-1} + (1 - \alpha) \cdot \hat{x}_{t-1} & \mbox{otherwise}
\end{cases}
\end{equation}

\begin{figure}
\centering
\includegraphics[width=0.85\textwidth]{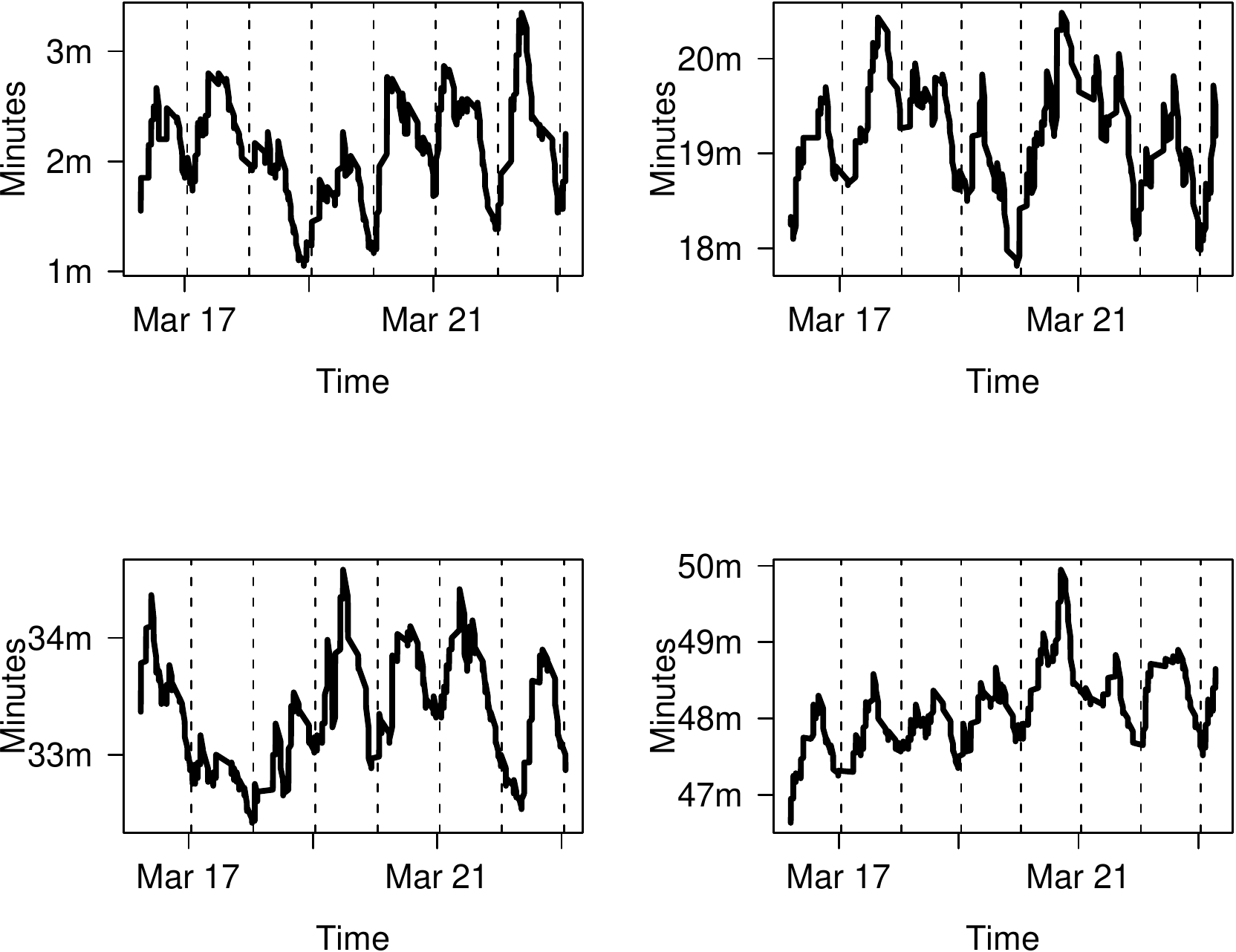}
\caption{Timing pattern of scanners connecting to our bridge. The $x$-axes depict the time in days
whereas the $y$-axes show the smoothed minute when scanners connected to the bridge.}
\label{fig:diurnal}
\end{figure}

We experimented with various smoothing factors $0 \le \alpha \le 1$ and found 0.05 to satisfyingly
illustrate the diurnal pattern. The result---a subset of the data ranging from March 16 to March
23---is shown in Figure \ref{fig:diurnal}. Each of the four diagrams represents one of the 15 minute
intervals. The diagrams show that depending on the time of the day, on average, scanners connect
either close to the respective 15 minute multiple or a little bit later.

We conjecture that the GFC maintains \emph{scanning queues}. When the DPI boxes discover a potential
Tor connection, the IP:port tuple of the suspected bridge is added to a queue. Every 15 minutes,
these queues are processed and all IP:port tuples in the queue are being scanned. In the evening,
more Chinese users try to access the Tor network than during the night. We assume that due to this
additional load, it takes a longer time for the GFC to process the queues.

\subsection{Blocking Malfunction}
During our experiments we noticed a sudden disappearence of active scanning and blocking. This lack
of scanning made it possible for us to successfully initiate Tor connections without causing bridges
to get scanned and blocked. A similar downtime was observed by Wilde \cite{tpo5}.

\begin{figure}
\centering
\includegraphics[width=0.9\textwidth]{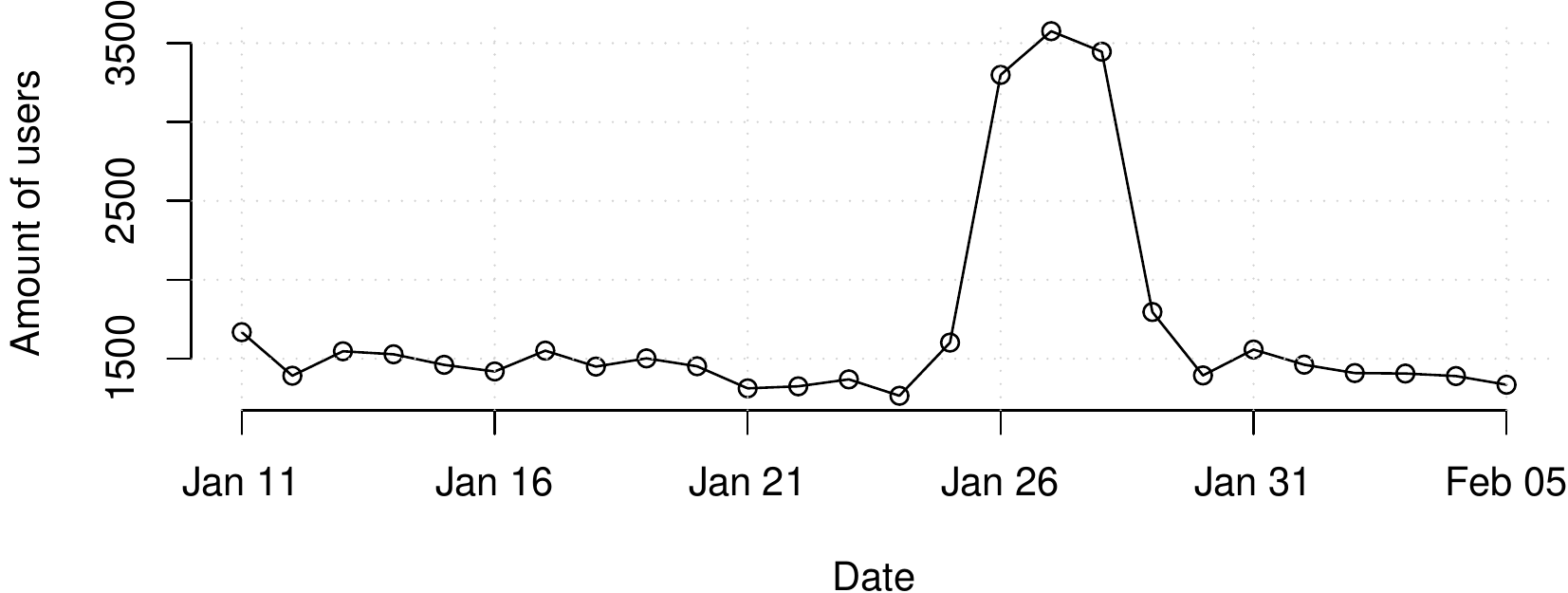}
\caption{Amount of Chinese bridge users between January 11 and February 5. The spike between January
25 and January 29 indicates a sudden increase of bridge users followed by a decrease. The reason for
the spike was an outage of the Chinese Tor blocking system.}
\label{fig:offline}
\end{figure}

The downtime started at around January 25 and lasted until January 28. During these
approximately three days many Chinese users were able to use the Tor network as can be seen in
Figure \ref{fig:offline}. The diagram shows a clear increase in usage followed by a drop to the same
level as before. After three days the blocking infrastructure was operating as before and Tor
connections from inside China were subject to the same active scanning as before the downtime.

We have no explanation for this downtime but it is worth pointing out that the downtime happened
during the Chinese new year celebration which lasted from January 23 to January 29. During this time
span the overall Internet usage in China decreased visibly as illustrated by Google's Transparency
Report \cite{trans}. This correlation could indicate that the downtime was not an accident but
rather a planned maintenance window.

\section{Evasion}
\label{sec:evasion}
The above sections explored the infrastructure and blocking mechanisms adopted by China to block
access to the Tor network. The question, which remains to be answered, is how to circumvent the
blocking. This section discusses different evasion techniques which greatly differ in their concept
but share the ultimate goal of circumventing the Chinese block.

\subsection{Obfsproxy}
\begin{figure}
\centering
\includegraphics[width=0.7\textwidth]{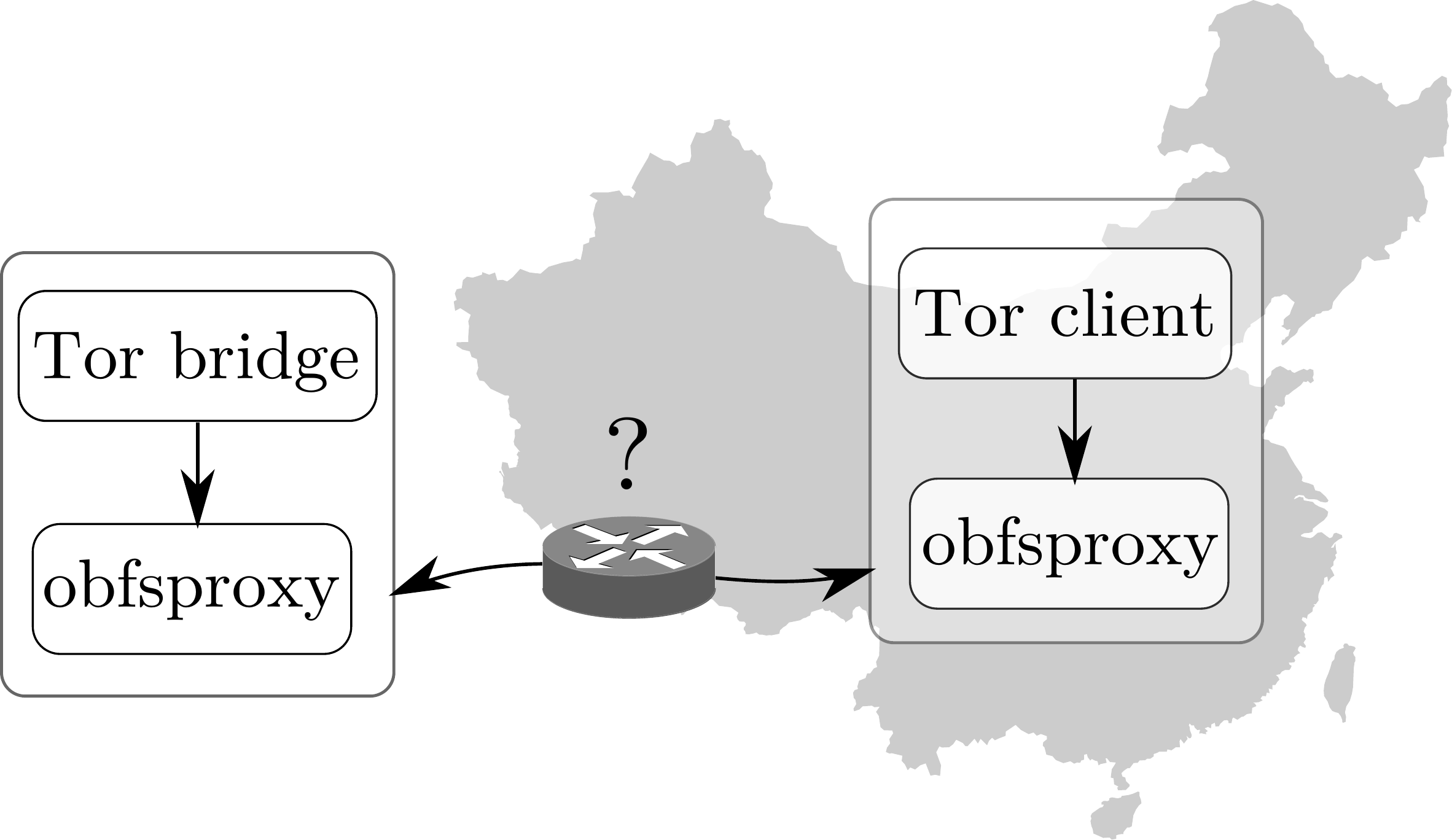}
\caption{Obfsproxy obfuscates the communication between a Tor client and a bridge. As a result, the
identification of Tor traffic will be much harder for DPI boxes.}
\label{fig:obfsproxy}
\end{figure}

The Tor Project is developing a tool called \emph{obfsproxy} \cite{obfsproxy}. The tool runs
independently of Tor and is obfuscating the network traffic it receives from the Tor process. An
example is shown in Figure \ref{fig:obfsproxy}. In the figure, two obfsproxy instances---a client
and a server---are exchanging obfuscated Tor traffic. Due to the obfuscation, the chinese DPI boxes
are not able to identify the TLS cipher list anymore.

Obfsproxy implements so called \emph{pluggable transport} meaning that the precise way of
obfuscation is determined by pluggable transport modules. As a result, one could implement a
transport module for HTTP whose purpose is to make Tor traffic look like HTTP. In
\cite{Moghaddam2012}, Moghaddam et al. describe the development of a pluggable transport module
which mimics Skype traffic.

As of March 24, the official obfsproxy bundle \cite{obfsproxy} contained a list of 13 hard-coded
obfsproxy bridges in its configuration file. From our VPS we tested the reachability of all of these
bridges by trying to connect to them via telnet. We found that not a single connection succeeded.
One bridge seemed to be offline and the connection to the remaining 12 bridges was aborted by
spoofed RST segments.

The above result raises the question whether the GFC is able to block all obfsproxy connections or
just the 13 hard-coded bridges. To answer this question, we set up a private obfsproxy bridge in
Sweden and connected to it from within China. We initiated several connections to it over
several hours and could always successfully establish a Tor circuit. We conclude that the IP:port
tuples of the 13 hard-coded obfsproxy bridges were added to a blacklist to prevent widespread use of
the official obfsproxy bundle. However, other obfsproxy bridges remain undetected by the GFC.

\subsection{Blacklisting the Blacklisters}
Chinese active scanners are almost indistinguishable from legitimate users. The scans originate from
seemingly random IP addresses and may even use the official Tor client to connect to suspected
bridges. Still, the scanners exhibit certain characteristics which make it possible to distinguish
between legitimate users and scanners.

\subsubsection{Deaf Time Window}
Figure \ref{fig:timing} indicates that scanners tend to connect at the beginning of every multiple
of 15 minutes. A bridge can exploit this knowledge by simply acting ``deaf'' against all new
connections from China for a while after a 15 minute interval has begun. Incoming TCP SYNs must be
silently dropped so that scanners are not able to connect.

Unfortunately, this strategy inevitably involves many false positives and leads to legitimate users
not being able to connect to the bridge.

\subsubsection{TCP SYN Retransmissions}
In December 2011, Pietrosanti \cite{fabio} observed that Chinese scanners retransmit only one TCP
SYN segment if the first segment is lost. This behaviour differs from modern Windows and Linux
machines which make use of more retransmissions. This phenomena can be equally exploited by
instructing bridges to deliberately ignore the first two TCP SYNs.

We implemented the above two blacklisting strategies in a software tool called ``brdgrd''. The
implementation offers the possibility to only accept the $n$th received SYN segment for a given
IP:port tuple. During our experiments with open SOCKS proxies we noticed that this technique is able
to shield off most scanners, but not all of them. We observed a few false positives which would
result in a bridge being blocked.

While the SYN retransmission technique is able to prevent a large fraction of scanners from
connecting, it causes collateral damage too since legitimate users can find themselves unable to
connect. This is particularly true if there is lots of packet loss between a user and a bridge.
Such a situation can effectively render a bridge unreachable for a user.

\subsection{Packet Fragmentation}
After all, the GFC is by its nature a distributed network intrusion detection system (NIDS). To
evade it, one can look through the glasses of an attacker and consider strategies outlined in a 14
year old paper written by Ptacek and Newsham \cite{Ptacek1998}.

One evasion technique described in the paper is \emph{packet fragmentation} which exploits the fact
that many NIDSes do not conduct packet reassembly. We used the tool \emph{fragroute}
\cite{fragroute} to enforce packet fragmentation on our machine in China. The tool allows the
interception, modification and rewriting of outgoing network traffic and implements most of the
attacks proposed in \cite{Ptacek1998}.

We configured fragroute to split the TCP byte stream to segments of 16 bytes each. At minimum this
configuration divides the 58 byte TLS cipher list into 4 TCP segments. In our test it took 5 TCP
segments to transmit the fragmented cipher list to our bridge as can be seen in Listing
\ref{lst:frag}.

\begin{lstlisting}[basicstyle=\footnotesize\ttfamily,caption={Fragmentation of the cipher list.},
label={lst:frag},captionpos=b]
 Packet n+0: c0 0a
 Packet n+1: c0 14 00 39 00 38 c0 0f c0 05 00 35 c0 07 c0 09
 Packet n+2: c0 11 c0 13 00 33 00 32 c0 0c c0 0e c0 02 c0 04
 Packet n+3: 00 04 00 05 00 2f c0 08 c0 12 00 16 00 13 c0 0d
 Packet n+4: c0 03 fe ff 00 0a 00 ff
\end{lstlisting}

Despite awaiting several multiples of 15 minutes and initiating several fragmented Tor connections,
we never noticed any active scanning of our bridge. This experiment indicates that the GFC does not
conduct packet reassembly. A similar observation was made by Park and Crandall \cite{Park2010}.

We conclude that packet fragmentation on the client side is sufficient to evade detection by the DPI
boxes. Still, this is an unpractical solution given that this method must be supported by \emph{all}
connecting Chinese users. A single user which does not use it triggers active scanning which then
leads to the block of all bridges, that the user is connecting to. Another disadvantage is the
significant protocol overhead due to the small TCP segments which leads to a decrease in throughput.

Due to the problems associated with client side fragmentation, we implemented a way of server side
fragmentation in brdgrd. We extended the tool to \emph{rewrite the TCP window size} announced by the
bridge to the client inside the SYN/ACK segment. Due to the rewritten and small window size, the
client does not transmit the complete TLS cipher list inside one segment but rather splits it across
two or more segments. In our tests we could successfully establish Tor connections from within China
without causing scanning. It must be noted that the rewriting is done without the bridge knowing and
is an abuse of the TCP protocol.

\subsection{Bridge Authorisation}
Scanners can be locked out if bridges are extended to provide a way of authorisation. Such a scheme
is developed by Smits et al. \cite{Smits2011}. The authors propose to use a single packet
authorisation mechanism. That way, a bridge can authorise a user and does not reveal its online
status before the authorisation was successful. For an unauthorised user, such as a scanner, a
bridge would appear to be offline.

\section{Conclusions}
\label{sec:conclusion}
China goes to great lengths to prevent its citizens from accessing the free Internet including the
Tor anonymity network. We showed how access to Tor is being denied and we conjectured how the
blocking and filtering infrastructure is designed. In addition, we discussed and proposed
countermeasures intended to ``unblock'' the Tor network.

Our findings include that the Great Firewall of China might spoof IP addresses for the purpose of
scanning Tor bridges and that domestic as well as ingress traffic does not seem to be subject to Tor
fingerprinting. We also showed that the firewall is easily circumvented by packet fragmentation.

Tor fingerprinting and active scanning is effective for the firewall because Tor traffic is
currently distinguishable from what is regarded as harmless traffic in China. Since Tor is being
used more and more as censorship circumvention tool, it is crucial that this distinguishability is
minimised.

\section*{Acknowledgments}
\label{sec:acks}
We want to thank Fabio Pietrosanti for helping with the experiments and Harald Lampesberger for
helpful comments. In addition, we want to thank the Tor developers (in particular George Kadianakis
and Arturo Filast\`{o}) for their valuable feedback and help.

Our raw data and code are available at: \\
\url{http://www.cs.kau.se/philwint/static/gfc/}.

\bibliographystyle{splncs}
\bibliography{literature}

\appendix
\section{Appendix}
\begin{lstlisting}[basicstyle=\scriptsize\ttfamily,captionpos=b,caption={Traceroute during scan.},label=lst:tr1]
traceroute to 60.208.176.112 (60.208.176.112), 30 hops max, 60 byte packets
 1  130.243.26.1 [AS1653]  0.656 ms  0.713 ms  0.759 ms
 2  193.10.220.173 [AS11908/AS1653]  0.662 ms  0.709 ms  0.722 ms
 3  193.10.220.178 [AS11908/AS1653]  0.527 ms  0.559 ms  0.567 ms
 4  193.11.0.145 [AS11908/AS1653]  0.592 ms  0.602 ms  0.629 ms
 5  130.242.85.17 [AS1653]  5.159 ms  5.179 ms  5.183 ms
 6  130.242.83.41 [AS1653]  5.186 ms  5.169 ms  5.173 ms
 7  109.105.102.13 [AS2603]  5.167 ms  5.272 ms  5.273 ms
 8  213.248.97.93 [AS1299]  5.338 ms  5.345 ms  5.353 ms
 9  80.91.247.206 [AS1299]  5.231 ms  5.238 ms  5.241 ms
10  213.155.130.175 [AS1299]  15.198 ms  15.143 ms  15.104 ms
11  213.248.64.34 [AS1299]  178.117 ms  108.001 ms  107.987 ms
12  80.91.254.177 [AS1299]  179.888 ms  180.010 ms  179.979 ms
13  213.248.71.90 [AS1299]  180.961 ms  180.973 ms  180.961 ms
14  219.158.30.229 [AS4837]  409.490 ms  409.363 ms  409.331 ms
15  219.158.96.197 [AS4837]  412.345 ms  412.534 ms  412.509 ms
16  219.158.3.61 [AS4134/AS4837]  407.295 ms  407.860 ms  407.381 ms
17  219.158.24.22 [AS4837]  448.474 ms  448.999 ms  448.994 ms
18  119.164.221.230 [AS4837]  450.969 ms  451.085 ms  451.091 ms
19  124.128.32.46 [AS4837]  395.287 ms  395.245 ms  395.233 ms
20  * * *
\end{lstlisting}

\begin{lstlisting}[basicstyle=\scriptsize\ttfamily,captionpos=b,caption={Traceroute after scan.},label=lst:tr2]
traceroute to 60.208.176.112 (60.208.176.112), 30 hops max, 60 byte packets
 1  130.243.26.1 [AS1653]  0.874 ms  0.877 ms  0.845 ms
 2  193.10.220.173 [AS11908/AS1653]  0.656 ms  0.670 ms  0.766 ms
 3  193.10.220.178 [AS11908/AS1653]  0.532 ms  0.566 ms  0.572 ms
 4  193.11.0.145 [AS11908/AS1653]  0.630 ms  0.640 ms  0.672 ms
 5  130.242.85.17 [AS1653]  5.182 ms  5.207 ms  5.219 ms
 6  130.242.83.41 [AS1653]  5.237 ms  5.178 ms  5.178 ms
 7  109.105.102.13 [AS2603]  13.578 ms  13.602 ms  13.608 ms
 8  213.248.97.93 [AS1299]  5.114 ms  5.126 ms  5.322 ms
 9  80.91.247.206 [AS1299]  5.319 ms  5.321 ms  5.310 ms
10  213.155.130.175 [AS1299]  15.046 ms  15.050 ms  15.038 ms
11  213.248.64.34 [AS1299]  152.373 ms  108.036 ms  108.045 ms
12  80.91.254.177 [AS1299]  179.995 ms  180.034 ms  180.039 ms
13  213.248.71.90 [AS1299]  180.160 ms  180.142 ms  180.268 ms
14  219.158.30.229 [AS4837]  410.041 ms  409.339 ms  409.286 ms
15  219.158.96.197 [AS4837]  412.971 ms  413.250 ms  413.192 ms
16  219.158.3.61 [AS4134/AS4837]  407.740 ms  407.721 ms  407.714 ms
17  219.158.24.22 [AS4837]  448.865 ms  448.850 ms  448.858 ms
18  119.164.221.230 [AS4837]  450.912 ms  451.291 ms  451.298 ms
19  124.128.32.46 [AS4837]  395.223 ms  395.154 ms  395.149 ms
20  60.208.176.112 [AS4837]  458.314 ms  458.502 ms  458.502 ms
\end{lstlisting}

\end{document}